# Role of Dirac nodal lines and strain on the high spin Hall conductivity of epitaxial IrO$_2$ thin films

Arnab Bose[1*], Jocienne N. Nelson[2], Xiyue S. Zhang[1], Rakshit Jain[2], D. G. Schlom[3,4], D. C. Ralph[2,4], D. A. Muller[1,4], K. M. Shen[2,4] and R. A. Buhrman[1]

*1 School of Applied and Engineering Physics, Cornell University, Ithaca, New York 14853, United States*
*2 Department of Physics, Cornell University, Ithaca, New York 14853, USA*
*3 Department of Materials Science and Engineering, Cornell University, Ithaca, New York 14853, US*
*4 Kavli Institute at Cornell for Nanoscale Science, Ithaca, New York 14853, United States*

\* ab2729@cornell.edu

**Abstract**
We report *spin-torque ferromagnetic resonance* (ST-FMR) studies of the efficiency ($\xi_{\text{DL}}$) of the damping-like spin-orbit torque exerted on an adjacent ferromagnet film by current flowing in epitaxial (001) and (110) IrO$_2$ thin films. Angle resolved photoemission spectroscopy studies have shown IrO$_2$ exhibits *Dirac nodal lines* (DNL) in the band structure, which could enable a very high spin Hall conductivity, $\sigma_{SH}$. The (001) films exhibit exceptionally high $\xi_{\text{DL}}$ ranging from 0.45 at 293 K to 0.65 at 30 K which sets the lower bounds of $\sigma_{SH}$ to be $1.9 \times 10^5 \Omega^{-1} \text{m}^{-1}$ and $3.75 \times 10^5 \Omega^{-1} \text{m}^{-1}$ respectively, ten times higher and of opposite sign than the theoretical prediction. We observe a substantial reduction of $\sigma_{SH}$ in anisotropically-strained (110) films, which suggests that the DNLs that are present in the (001) films and contribute to $\sigma_{\text{SH}}$, are disrupted and gapped due to the large anisotropic strain in (110) films, which in turn significantly lowers $\xi_{\text{DL}}$.

Since the discovery of a "giant" spin Hall effect (SHE)[1] in certain heavy metal elements there has been an intense effort to identify and develop new and technologically viable, heavy-metal-based thin film materials that could generate spin currents with even greater efficiency to exert spin-orbit torques (SOT)[2] on adjacent ferromagnetic nanostructures. In parallel, there have been wide ranging fundamental studies of the spin currents that can arise from robust, intrinsic spin-orbit interaction (SOI) effects in more exotic systems including topological insulators[3,4], transition metal dichalcogenides with broken crystalline symmetry[5], Weyl[6] and Dirac semimetals[7] where gapless electronic excitations are protected by topology and symmetry. Here we experimentally study strong SOT from the topological semimetal IrO$_2$ in (001) and (110) normal films, which exhibit distinctly different SHE strengths. The high resistivity (001) films have a quite high spin Hall conductivity (SHC) with behavior indicative of being in the "*dirty metal regime*", while the anisotropically strained (110) films have a substantially smaller SHC and are in the "*clean metal regime*".

A recent focus of SOT research has been the semimetallic SrIrO$_3$ where theoretical calculations and ARPES indicate that the tilt of the IrO$_6$ octahedra in the orthorhombic phase leads to narrow $t_{2g}$ conduction bands with near-degeneracies and band crossings protected by the nonsymmorphic crystal symmetry[8,9,10]. SOT measurements on SrIrO$_3$/FM bilayers show a large effective-SHC $\sigma_{\text{SH}}^{\text{eff}} \sim 10^5 \Omega^{-1} \text{m}^{-1}$ [11,12,13,14]. Here $\sigma_{\text{SH}}^{\text{eff}} \equiv T_{\text{int}} \sigma_{\text{SH}}$, where $T_{\text{int}}(<1)$ is the interfacial spin transparency and $\sigma_{\text{SH}}$ is the actual SHC of the material. It is intriguing that Ref 11 and 12 report *opposite signs* of $\xi_{\text{DL}}$ and also, a significant in-plane crystalline anisotropy is reported for $\xi_{\text{DL}}$ in Ref. 11,13, but not in Ref. 12. The simplest iridate is IrO$_2$ which in the rutile phase consists of chains of edge-sharing IrO$_6$ octahedra running along the *c*-axis with corner-sharing connections among counter-rotated neighboring chains in the (001) plane. Due to the mixture of both edge and corner sharing $IrO_6$ octahedra, and a relatively direct coordination between Ir atoms that provides an additional hopping channel, the overall result is increased *p-d* orbital hybridization, which broadens the $t_{2g}$ conduction band compared to other, corner-sharing, iridates[9,10]. This reduced electron correlation, which also is found in tetragonal $SrIrO_3$ where neighboring octahedra are not counter-rotated and similarly results in larger bandwidths, might suggest that $IrO_2$ would be a material with a relatively low SHC, as is reported for tetragonal $SrIrO_3$[11]. Recent *ab initio* density functional theory (DFT) calculations, however, indicate that the $IrO_2$ band-structure contains two types of Dirac nodal lines (DNLs) associated with the nonsymmorphic crystal symmetry and the rutile structure[15]. Type-1 DNLs allow an anticrossing near the Fermi surface due to SOI and can therefore contribute a large SHC with a maximum value of $-2.5 \times 10^4 (\hbar/2e) \Omega^{-1} m^{-1}$ [15], roughly in accord with the results of an early inverse-SHE study of polycrystalline/amorphous $IrO_2$ films, although of opposite sign [16]. Subsequent ARPES studies confirmed the existence of these DNLs in epitaxial $IrO_2$ thin films and bulk crystals[17,18].

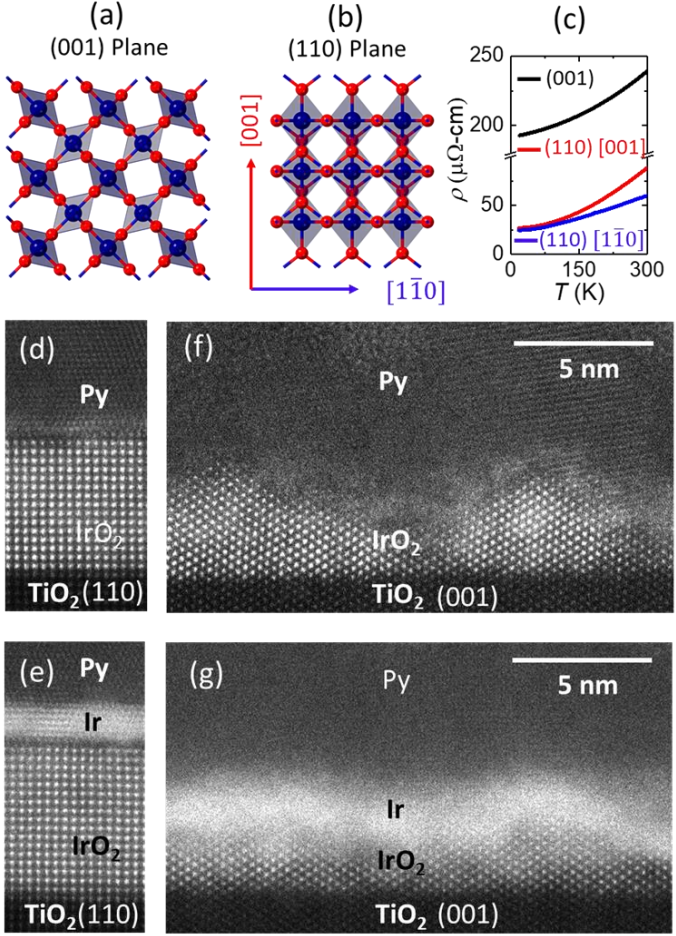

**Figure 1.** (a), (b) Schematic top-down view representation of the (001) and (110) plane of $IrO_2$ illustrating the edge to corner connectivity of the octahedra in the (001) plane as opposed to the corner to corner and edge to edge connectivity in the (110) plane. (c) Resistivity of (001) and (110) $IrO_2$ films as a function of temperature and for the two different principal axes $[1\bar{1}0]$ and $[001]$ of the (110) film. (d-g) Annular dark field STEM image of (110) $IrO_2$/Py, (110) $IrO_2$/Ir/Py, (001) $IrO_2$/Py and (001) $IrO_2$/Ir/Py respectively.

Here we report values of $\sigma_{SH}^{eff}$ for (001) and (110) $IrO_2$ films grown epitaxially on $TiO_2$ rutile substrates as determined by ST-FMR measurements of the efficiency of the damping-like $\xi_{DL}$ and field-like $\xi_{FL}$ spin torques, exerted on an adjacent $Ni_{80}Fe_{20}$ (Py) thin film. $\sigma_{SH}^{eff}$ is then obtained from $\sigma_{SH}^{eff} = \xi_{DL}\sigma_e$, where $\sigma_e$ is the film's electrical conductivity. Due to the thermodynamically favored reactions between Fe and $IrO_2$[19] and the unavoidable presence of undercoordinated O ions at the surfaces of a rutile film[20], the Fe in the Py near the interface becomes oxidized if deposited directly on the $IrO_2$. We therefore inserted a thin Ir passivation layer (~1 nm) between the $IrO_2$ and the Py to avoid this oxidation (Figure 1d-g). (We have determined that Ir has a minimal SHE, $\xi_{DL} \leq 0.02$.) Even with the significant spin current attenuation arising from this Ir layer, at room temperature (RT) we determined $\xi_{DL}$ to be 0.45 for current flow in the plane of the low $\sigma_e$ (001) films, rising to 0.65 at 30K, corresponding to $\sigma_{SH}^{eff} = 1.9 \times 10^5 (\hbar/2e)\Omega^{-1}m^{-1}$ at RT, and $3.75 \times 10^5 (\hbar/2e)\Omega^{-1}m^{-1}$ at 30 K. These lower bounds are more than an order of magnitude higher, and of opposite sign, than the available theoretical prediction for $\sigma_{SH}$ [15]. For the (110) $IrO_2$ films, which are anisotropically-strained, the SHC is substantially reduced while $\sigma_e$ is increased (Figure 1c), and both quantities are slightly anisotropic. At RT $\sigma_{SH}^{eff}$ varies from $7.8 \times 10^4 (\hbar/2e)\Omega^{-1}m^{-1}$ to $8.8 \times 10^4 (\hbar/2e)\Omega^{-1}m^{-1}$ between the lower conductivity [001] and higher conductivity $[1\bar{1}0]$ directions. The anisotropy in $\sigma_{SH}^{eff}$ increases as $\sigma_e$ increases with decreasing temperature (*T*), with $\sigma_{SH}^{eff}$ becoming $9.6 \times 10^4 \Omega^{-1}m^{-1}$ and $1.85 \times 10^5 \Omega^{-1}m^{-1}$, respectively, at 30 K. We attribute the large reduction of $\sigma_{SH}^{eff}$ in the anisotropically-strained (110) IrO$_2$ compared to isotropically-strained (100) IrO$_2$ as being due to disruption of the type-1 DNLs by the anisotropic strain.

For this study, we grew IrO$_2$ thin films of 5 nm thickness by reactive oxide molecular beam epitaxy (MBE) on (001) and (110) single crystal TiO$_2$ substrates at room temperature, as described in Supporting Information[21] and in Ref. [9,10,22,23]. The IrO$_2$ grows centrosymmetrically on (001) TiO$_2$, with a coherent, isotropic in-plane biaxial tensile strain of $\epsilon_{11} = \epsilon_{22} \approx 2\%$. The nonsymmorphic symmetry is preserved under this isotropic strain. In contrast the (110) films grow with a compressive strain (~ -5 %) in the [001] direction and a tensile strain (~ 2%) in the [$1\bar{1}0$] direction films due to the lattice mismatch between IrO$_2$ and (110) TiO$_2$. It has been found that this anisotropic strain for (110) IrO$_2$ films persists up to at least a 16 nm thickness[10]. This strain breaks the nonsymmorphic symmetry of the (110) films.

The *T*-dependent resistivities of the different films and for different current directions in the (110) case are shown in Figure 1c. The resistivity of the (001) film, which is independent of the direction of current flow, is high and exhibits only a small residual resistivity ratio (RRR). The (110) films have a lower resistivity and higher RRR, indicative of a cleaner metal, with the resistivity being dependent upon direction of the current flow. This anisotropy does not, however, appear to be due primarily to the anisotropic strain in these films since the nearly 50% difference in resistivity at RT between current flow in the [001] and [$1\bar{1}0$] directions, Figure 1c, is roughly consistent with the 77% difference in resistivity at RT in unstrained IrO$_2$ single crystals[24,22].

For the ST-FMR measurements[25,26,27] four series of multilayers were prepared as described in Ref [21]. Series 1: (001)IrO$_2$(5nm)/Ir (1nm)/Py(3-6nm)/TaO$_x$(1.5nm); Series 2: (110)IrO$_2$(5nm)/Ir(1)nm)/Py(3-6nm)/TaO$_x$(1.5nm); Series 3: (001)IrO$_2$(5nm)/Py(3-7nm)/TaO$_x$(l.5nm); and Series 4: (110)IrO$_2$(5nm)/Py(2.7-7nm)/TaO$_x$(1.5nm). The multilayers with the Ir passivation layer, Series 1 and 2, were the major focus of the SOT study, while Series 3 and 4 were used to quantify the benefit and cost of the use of the Ir spacer.

Cross-sections of examples of each of the four types of IrO$_2$ multilayer were examined by high resolution scanning transmission electron microscopy (STEM) and EELS. Annular dark field STEM images of the cross-sections are shown in Figure 1d-g. The epitaxial (110) films are atomically flat on the lateral scale of the cross-sectional image while the Py is polycrystalline, both with and without the textured Ir spacer (Figure 1d-e). The (001) IrO$_2$ films are quite faceted (Figure 1f-g), consistent with the high surface energy of the (001) rutile surface[28]. The faceted IrO$_2$ (001) surface is most clearly seen in the sample without the Ir spacer (Figure 1f) as the latter wets the surfaces of the IrO$_2$ facets, obscuring the dark-field contrast with the Py. Spatially resolved EELS measurements show that the Ir is quite effective in passivating the surface of the IrO$_2$ for both the (110) and the faceted (001) films, with there being no signal of oxidized Fe atoms[21]. For the samples without the Ir spacer, the EELS data indicate significant Fe oxidation in the vicinity of the IrO$_2$ surface for the faceted (001) films. There is less Fe oxide at the unpassivated (110) IrO$_2$ surface, consistent with the lower density of under-coordinated O ions at the (110) rutile surface.

The details of the ST-FMR technique we used are provided in the Ref [21]. In brief we measure the FMR resonant response to a microwave current applied to the HM/FM coplanar waveguide in the presence of an in-plane magnetic field. This response consists of a symmetric Lorentzian voltage signal of amplitude *S* centered about the resonant field $H_o$ due to the in-plane damping-like torque, and an antisymmetric Lorentzian response, of amplitude *A*, to the out-of-plane torque resulting from the microwave Oersted field generated by the current flowing in the HM, plus any additional interfacial field-like torque contribution from the incident spin current. The result can be expressed in term of a quantity $\xi_{FMR}$[26] where

$$\frac{1}{\xi_{FMR}} \equiv \frac{\frac{A}{S}\left(\frac{\hbar}{e}\right)}{\left(\mu_0 M_s d_{Py} d_{IrO_2} \sqrt{1+\left(\frac{M_{eff}}{H_0}\right)}\right)} = \frac{1}{\xi_{DL}}\left(1 + \frac{\hbar}{e}\frac{\xi_{FL}}{\mu_0 M_s d_{Py} d_{IrO_2}}\right) \quad \text{Eqn. 1.}$$

Here $d_{Py}$ and $d_{IrO_2}$ are the thicknesses of the Py and IrO$_2$ layers, $M_s$ is the saturation magnetization of the Py as measured by vibrating sample magnetometry and $M_{eff}$ is the Py's out-of-plane demagnetization field.

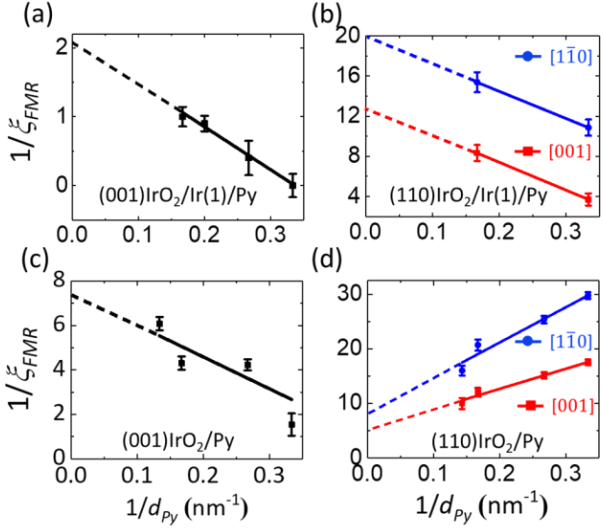

**Figure 2.** The inverse of the ST-FMR apparent spin torque efficiency (1/$\xi_{FMR}$) at room temperature as a function of the inverse of Py thickness (1/$d_{Py}$) for the different series of samples: (a) (001)IrO$_2$/Ir(1)/Py; (b), (110)IrO$_2$/Ir(1)/Py for both the [1$\bar{1}$0] (red color) and [001] directions (black color); (c) (001)IrO$_2$/Py; and (d) (110)IrO$_2$/Py.

In Figure 2, we show plots of 1/$\xi_{FMR}$ vs. 1/$d_{Py}$ as obtained at RT for all four series of samples. The 1/$d_{Py}$ = 0 intercepts of the linear fits to the variation of 1/$\xi_{FMR}$ provide 1/$\xi_{DL}$ while $\xi_{FL}$ is determined from the slope of the linear fit to Eqn. 1. The results are presented in Table I. The high resistivity (001) IrO$_2$ with the 1 nm passivation layer exhibits an exceptionally high, $\xi_{DL}$ = 0.45±0.03 (Figure 2a), notwithstanding the spin current attenuation of the Ir spacer. The lower resistivity (110) films with the Ir spacer have lower $\xi_{DL}$ along the [1$\bar{1}$0] and (001) directions with a smaller anisotropy in $\sigma_{SH}^{eff}$ as shown in Table 1.

When the Ir spacer layer was not utilized for the (001) IrO$_2$ films the result was more than a factor of three degradation of $\xi_{DL}$ from the Ir passivated result (Figure 2c-d), which we attribute to the substantial oxidation of the interfacial iron in the Py layer, as revealed by the EELS study[26]. In contrast, in the absence of the Ir spacer for the (110) samples $\xi_{DL}$ is approximately doubled in amplitude for both the [1$\bar{1}$0] and [001] directions, and with a sign reversal of $\xi_{FL}$. We conclude that the degree of interfacial Fe oxidation indicated by EELS for the atomically smooth (110) films is not very detrimental, if at all, to interfacial spin transport. From the (110) results we estimate the Ir spin diffusion length $\lambda_s$ to be ~ 1.2 nm, consistent with a recent work[29]. This results in the attenuation of the spin current entering the Ir layer from the IrO$_2$ by approximately 50% before it impinges on the Py.

We measured the SOTs of the Ir passivated samples from RT to 30 K, with the results shown in Figure 3. For the high-resistivity (001) IrO$_2$ films both $\xi_{DL}$ and $\xi_{FL}$ increased substantially as $T$ was decreased, with $\xi_{DL}$ reaching 0.65±0.07 at 30 K, and with the proportionate increase in $\xi_{FL}$ being even larger (Figure 3a). For both the [1$\bar{1}$0] and [001] directions of the (110) films $\xi_{DL}$ decreased with decreasing $T$, with the percentage decrease being the larger for the lower resistivity [1$\bar{1}$0] direction, while the smaller $\xi_{FL}$ increased substantially for the [1$\bar{1}$0] direction (Figure 3b) and was more or less constant for the higher resistivity [001] direction (Figure 3c). The quite variable behavior of $\xi_{FL}$ is suggestive that the role of the interface is more complex than, for example, is the case for typical Pt/FM system.

| Parameters | (001)/Ir/Py | (001)/Py | (110)[1$\bar{1}$0]/Ir/Py | (110)[1$\bar{1}$0]/Py | (110)[001]/Ir/Py | (110)[001]/Py |
|---|---|---|---|---|---|---|
| $\xi_{DL}$ | 0.45±0.03 | 0.14±0.01 | 0.05±0.008 | 0.12±0.01 | 0.08±0.01 | 0.18±0.01 |
| $\xi_{FL}$ | -0.05±0.005 | -0.01±0.005 | -0.02±0.005 | 0.04±0.005 | -0.02±0.005 | 0.02±0.005 |
| $\sigma_{SHE}^{eff}$ ($10^5 \hbar/2e\ \Omega^{-1}m^{-1}$) | 1.9±0.15 | 0.57±0.05 | 0.78±0.11 | 1.95±0.2 | 0.88±0.12 | 2.13±0.2 |

**Table 1:** Values for the damping-like and field-like spin torque efficiencies, as measured at room temperature by ST-FMR, for 5 nm epitaxial IrO$_2$ films with and without a 1 nm Ir passivation layer between the IrO$_2$ and the Py spin current detector layer.

We have used the results presented in Figure 3a-c along with the measured variation of the resistivity (Figure 1c) to obtain the variation of $\sigma_{SH}^{eff}$ for the two different Ir-passivated IrO$_2$ surfaces, and for the two principal directions of the (110) samples, both as a function of $T$ and $\sigma_e$. Results are shown in. Figure 3d-e. The $T$-dependent resistivity indicates that that IrO$_2$ has a metallic behavior (Figure 1c) with electron transport through the $t_{2g}$ conduction band. For the intrinsic SHE it is expected that $\sigma_{SH}$ should be more or less independent of $\sigma_e$ in the "*clean metal*" regime where carrier lifetime is not a factor in determining the overall spin Berry curvature (SBC) and hence $\sigma_{SH}$[30]. However as $\sigma_e$ decreases, that is as the carrier lifetime becomes shorter, $\sigma_{SH}$ is expected to degrade, at first gradually and then more rapidly as $\sigma_e$ decreases further [30]. Previous studies have examined the transition between the dirty to clean metal regimes of the SHE in Pt-based alloys and composites[31,32] and in epitaxial SrRuO$_3$ thin films[33] which are compared with the IrO$_2$ films by plotting $\sigma_{SH}^{eff}$ vs $\sigma_e$ (Figure 3e). (We point out that these plots are only qualitative comparisons that do not account for the differences in carrier density and band structure between the different materials.)

The nearly constant $\sigma_{SH}^{eff}$ with varying $\sigma_e$ (temperature) for the (110) IrO$_2$ film with current flow in the low resistivity [1$\bar{1}$0] direction suggests that in this case the material is in or near the "*clean metal regime*", but with a lower value of $\sigma_{SH}^{eff}$, with the behavior for the [001] direction showing a somewhat greater variation with $\sigma_e$ and larger amplitude overall (Figure 3e). The (001) film appears to be well within the "*dirty metal regime*" but still reaches a quite high maximum value of $\sigma_{SH}^{eff}$, $3.75 \times 10^5 (\hbar/2e)\Omega^{-1}m^{-1}$ (Figure 3e). The data suggests that $\sigma_{SH}^{eff}$ of the (001) film, which is linearly increasing with its $\sigma_e$, would further increase and saturate at an even higher value than measured here if a sufficiently higher quality, lower defect density (001) film could be produced with a high enough $\sigma_e$ to be near or in the clean metal regime. We also note that since the variations in $\sigma_{SH}^{eff}$ between the different types of epitaxial IrO$_2$ films and the different current directions of the (110) films do not fall along a common scaling curve as a function of $\sigma_e$ ($\sigma_{SH}^{eff} = \theta_{SH}\sigma_e$), it appears that the primary cause of the difference is not due to the difference in carrier lifetime.

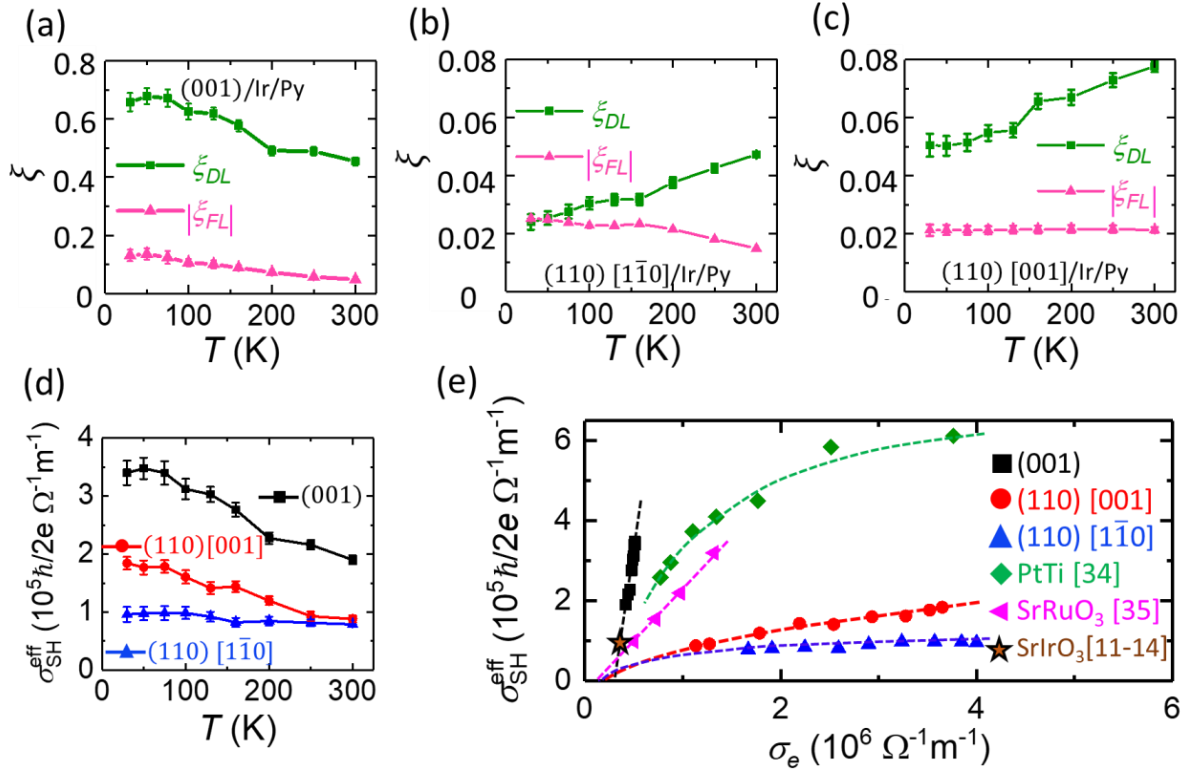

**Figure 3.** Temperature dependence of $\xi_{DL}$ and $|\xi_{FL}|$ of the Ir passivated samples for the (001) IrO$_2$/Ir(1)/Py series (a), (110) IrO$_2$/Ir(1)/Py series in the low resistivity [1$\bar{1}$0] direction (b) and (110) series along high resistive [001] direction (c). (d) $\sigma_{SH}^{eff}$ as a function of temperature for three different cases. (e) $\sigma_{SH}^{eff}$ as a function of electrical conductivity ($\sigma_e$) of the 5 nm IrO$_2$ films for the three different cases. Also shown for qualitative comparison are published results of $\sigma_{SH}^{eff}$ vs. $\sigma_e$ for SrRuO$_3$, SrIrO$_3$, and 6 nm Pt films where variable numbers of sub-monolayer insertions of Ti were employed to vary $\sigma_e$.

In seeking a basic understanding of these results we have the challenge that, while the DFT band structure calculations for IrO$_2$ have generally been experimentally confirmed by ARPES measurements on IrO$_2$ films, within the accuracy of that technique, although there are discernable differences between measurement and calculation for the anisotropically strained (110) films up to a film thickness of 16 nm,[10,17] we have measured the SHC to be ~10 times the prediction and of opposite sign. (We comment that theoretical underestimation of the SHC is also an issue in simpler spin Hall materials such as Pt and Pt alloys[31,32].) While keeping that disagreement in mind we note that the band structure calculations[15] and ARPES measurements show[17], as mentioned above, that the IrO$_2$ rutile crystal structure possesses two types of DNLs. The DFT calculations predict that the SOC induced anticrossing of bands forming type-1 DNLs near the Fermi energy can provide a large contribution to the SHC. We speculate that the band structure distortion due to the anisotropic strain in the (110) case, and the resulting breaking of the nonsymmorphic crystal symmetry, act to gap the DNLs and suppress the SHC. (A bandgap induced by interactions other than SOC will reduce the SBC.) The distinctly different values and behavior for $\sigma_{SH}^{eff}$ between the (001) and (110) films, correlated with the absence and presence of anisotropic strain in the two cases, strongly support the role of DNLs, and the effect of strain on them, in determining the amplitude of the SHC, notwithstanding the disagreement between theoretical prediction and experimental results with respect to both amplitude and sign. Along this line of argument, it also seems reasonable to attribute the smaller variation of the SHC between current flow in [$1\bar{1}0$] and [001] directions of the (110) plane to the modulation of the SHC by the anisotropic strain in those films. We further note the stark difference between SrIrO$_3$ and IrO$_2$ as the (001) IrO$_2$ film exhibits a three times stronger (six times after correcting spin attenuation though the Ir spacer) SHC than SrIrO$_3$ which requires octahedral rotation and strong electron correlation to maximize its SHC. It appears that strong electron correlation and a narrow conduction band, by itself, is not essential for obtaining a very high SHC.

In summary, we have measured a very strong, orientation-dependent SHE in epitaxial (001) and anisotropically-strained (110) IrO$_2$ thin films with $\sigma_{SH}$ ten times higher and opposite in sign compared to the prediction. Even with the insertion of a 1 nm thick Ir passivation layer, which significantly attenuates the spin current, $\xi_{DL}$ is 0.45±0.03 at RT and 0.65 at 30 K for the higher-resistivity, isotropically strained (001) films. Accounting for the spin attenuation of the 1 nm Ir spacer and assuming $T_{int}$ of approximately 50%, the $\sigma_{sh}$ would be increased by factor of four. For the (110) films $\sigma_{SH}^{eff}$ is lower, while still well above the predicted values for IrO$_2$, and is anisotropic between current flow in the [$1\bar{1}0$] and [001] directions. We tentatively attribute the lower $\xi_{DL}$ (lower $\sigma_{SH}^{eff}$) of the anisotropically strained (110) films, compared to that of the (001) films, to a gapping of the type-1 DNL due to band structure distortion. These results support the theoretical conclusion that the DNLs play a significant role in the large SHE in IrO$_2$, despite the quantitative disagreement with the magnitude and sign of the experimental results. The (001) film results, combined with the ease of the room temperature growth of IrO$_2$ epitaxial films, are very promising for application. Achieving a quantitatively correct understanding of the extraordinarily large SHE in a material with strong DNL but a broad $t_{2g}$ conduction band appears to be a continuing theoretical challenge.

## Supporting Information

Additional details of the growth and characterization of IrO$_2$ thin films, sample preparation and ST-FMR measurements and analysis can be found in the Supporting Information.

## Author contributions

## Acknowledgments

We thank Ryan Tapping for providing data on the spin torque efficiency of Ir films. X.S. Zhang thanks Celesta Chang and M. Thomas for assistance with the STEM/EELS studies. Funding: This work was supported in part by the National Science Foundation's MRSEC program (DMR-1719875) through the Cornell Center for Materials Research and the NSF's Platform for the Accelerated Realization, Analysis, and Discovery of Interface Materials (PARADIM)) under Cooperative Agreement No. DMR-1539918, by the Office of Naval Research (N00014-19-1-2143), by NSF (DMR-1709255), and by the NSF (ECCS-1542081) through use of the Cornell Nanofabrication Facility/National Nanotechnology Coordinated Infrastructure. R.J. was supported by the US Department of Energy (DE-SC0017671).